\documentclass[a4paper,12pt]{article}
\usepackage{amssymb}
\usepackage{mathrsfs}
\usepackage{bbm}
\usepackage{epsf,amsmath,amssymb}
\usepackage[dvips,usenames]{color}
\usepackage{graphicx}
\usepackage{color}
\usepackage{colortbl}
\usepackage{epsfig}

\newlength{\dinwidth}
\newlength{\dinmargin}
\def \eff{{\text{eff}}}

\def\c{C}
\def\cs{{\c_7}}
\def\cn{{\c_9}}
\def\ct{{\c_{10}^{\rm eff}}}
\def\cne{\cn^{\rm eff}}
\def\cse{\cs^{\rm eff}}

\def\gl{\Gamma}

\def\l{\ell}

\def\d{{\rm d}}

\def\mh{\hat{m}}
\def\mbh{\mh_b}

\def\mvh{\mh_{\phi}}

\def\mlh{\mh_\l}

\def\sh{\hat{s}}

\def\a{{\cal A}}

\def\uh{{\hat{u}}}
\def\la{{\lambda}}

\def\be{\begin{equation}}
\def\ee{\end{equation}}
\def\ba{\begin{eqnarray}}
\def\ea{\end{eqnarray}}

\begin{document}

\title{\bf Probe a family non-universal $Z^{\prime}$ boson effects in $\bar{B}_s \to \phi\mu^+\mu^-$ decay}
\author{Qin Chang$^{a,c}$\footnote{Corresponding author. changqin@htu.cn}, Yin-Hao Gao$^{b}$\\
{ $^a$\small Department of Physics, Henan Normal University,
Xinxiang, Henan 453007, P.~R. China}\\
{$^{b}$\small Henan Institute of Science and Technology, Xinxiang 453003, P. R. China}\\
{$^{c}$\small Institute of Particle Physics, Huazhong Normal
University, Wuhan,
Hubei 430079, P. R. China}}

\date{}
\maketitle
\bigskip\bigskip
\maketitle \vspace{-1.5cm}

\begin{abstract}
Motivated by the recent measurement on ${\cal B}(\bar{B}_s\to \phi \mu^+\mu^-)$ by CDF collaboration, we study the effects of a family non-universal $Z^{\prime}$ boson on rare semileptonic $\bar{B}_s \to \phi\mu^+\mu^-$ decay. In our evaluations, we analyze the dependences of the dimuon invariant mass spectrum and normalized forward-backward asymmetry on $Z^{\prime}$ couplings and show that these observables are highly sensitive to new $Z^{\prime}$ contributions. Three limiting scenarios are presented in the detailed analyses. Numerically, within the allowed ranges of $Z^{\prime}$ couplings under the constraints from $\bar{B}_s-B_s$ mixing, $B\to\pi K$, $\bar{B}_d\to(X_s,K,K^{\ast})\mu^+\mu^-$ decays and so on, ${\cal B}(\bar{B}_s\to \phi \mu^+\mu^-)$ and $A_{FB}^{(L)}(\bar{B}_s\to \phi \mu^+\mu^-)$ could be enhanced by about $96\%$ and $17\%\,(133\%)$ respectively at most by $Z^{\prime}$ contributions. However,  ${\cal B}(\bar{B}_s\to \phi \mu^+\mu^-)$ is hardly to be reduced. Furthermore, the zero crossing in $A_{FB}(\bar{B}_s\to \phi \mu^+\mu^-)$ spectrum at low dimuon mass always exists.
\end{abstract}


\newpage

\section{Introduction}
Rare B decays induced by the flavor-changing neutral current~(FCNC) occur at loop level in the Standard Model~(SM) and thus proceed at a low rate.
They can provide useful information on the parameters of the SM and test its predictions. Meanwhile, they offer a valuable possibility of an indirect search of
new physics~(NP) for their sensitivity to the gauge structure and new contributions. Experimentally, the fruitful running of BABAR, Belle and Tevatron in the past decade provides a very fertile ground for testing SM and probing possible NP effects. As particle physics is entering the era of LHC, $B_s$ physics has attracted much more attention.

Recently, CDF collaboration has reported the first observation of the rare semileptonic $\bar{B}_s \to \phi\mu^+\mu^-$ decay and measured its branching fraction to be~\cite{CDFPhill}
\begin{equation} \label{eq:Heff}
{\cal B}(\bar{B}_s \to \phi\mu^+\mu^-)=[1.44\pm0.33({\rm stat.})\pm0.46({\rm syst.})]\times10^{-6}\,\quad {\rm CDF\,.}
\end{equation}
Theoretically, many evaluations for $\bar{B}_s \to \phi\mu^+\mu^-$ decay have been done within both SM and various NP scenarios (for example, Refs.~\cite{Erkol,Bsphill}). The SM prediction for ${\cal B}^{SM}(\bar{B}_s \to \phi\mu^+\mu^-)$~($\sim1.65\times10^{-6}$(QCDSR)~\cite{Erkol}, for example) agrees well with CDF measurement$~(1.44\pm0.57)\times10^{-6}$ for large experimental error. If more exact measurement on $\bar{B}_s \to \phi\mu^+\mu^-$ is gotten by the running  LHC-b and future super-B, the possible NP space will be strongly constrained or excluded. So, it is worth evaluating the effects of the possible NP, such as a family non-universal $Z^{\prime}$ boson, on $\bar{B}_s \to \phi\mu^+\mu^-$ decay.

A new family non-universal $Z^\prime$ boson could be naturally derived in certain
string constructions~\cite{string}, $E_6$ models~\cite{E6} and so
on. Searching for such an extra $Z^{\prime}$ boson is an important
mission in the experimental programs of Tevatron~\cite{Tevatron} and
LHC~\cite{LHC}. The general framework for non-universal $Z^{\prime}$ model has been developed in Ref.~\cite{Langacker}. Within such model, FCNC in $b\to s$ and $d$ transitions could be induced by family non-universal $U(1)^\prime$ gauge symmetries at tree level. Its effects on $b\to s$ transition have attracted much more attention and been widely studied. Interestingly, the behavior of a family non-universal $Z^\prime$ boson is helpful to resolve many puzzles in $B_{(u,d,s)}$ decays, such as ``$\pi K$ puzzle''~\cite{Barger1,Chang1}, anomalous $\bar{B}_s-B_s$ mixing phase~\cite{Liu,Chang2} and mismatch in $A_{FB}(B\to K^{\ast}\mu^{+}\mu^{-})$ spectrum at low $q^2$ region~\cite{Chang3,CDLv}.

Within a family non-universal $Z^\prime$ model, $\bar{B}_s \to \phi\mu^+\mu^-$ decay involves $b-s-Z^{\prime}$ and $\mu-\mu-Z^{\prime}$ couplings, which have been strictly bounded by the constraints from $\bar{B}_s-B_s$ mixing, $B\to\pi K^{(\ast)}$, $\rho K$, $\bar{B}_d\to X_s\mu\mu$, $K^{(\ast)}\mu\mu$ decays and so on~\cite{Chang1,Chang2,Chang3}.
So, it is worth evaluating the effects of a non-universal $Z^\prime$ boson on $\bar{B_s}\to\phi\mu^+\mu^-$ decay and checking whether such settled values of $Z^\prime$ couplings are permitted by CDF measurement on ${\cal B}(\bar{B}_s \to \phi\mu^+\mu^-)$.

Our paper is organized as follows. In Section~2,
we briefly review the theoretical framework for $b\to s l^+ l^-$ decay within both SM and a family
non-universal $Z^{\prime}$ model. In Section~3, the effects of
a non-universal $Z^{\prime}$ boson on $\bar{B}_s \to \phi\mu^+\mu^-$ decay are investigated in detail. Our conclusions are summarized in
Section~4. Appendix~A and B include all of the theoretical input
parameters.

\section{The theoretical framework for $b\to s l^+ l^-$ decays}\label{theo}
In the SM, neglecting the doubly Cabibbo-suppressed contributions, the effective Hamiltonian governing semileptonic $b\to s \ell^+\ell^-$ transition is given by~\cite{Altmannshofer:2008dz,Chetyrkin:1996vx}
\begin{equation} \label{eq:Heff}
{\cal H}_{\eff} = - \frac{4\,G_F}{\sqrt{2}}V_{tb}V_{ts}^{\ast}
\sum_{i=1}^{10} C_i(\mu) O_i(\mu) \,.
\end{equation}
Here we choose the operator basis given by Ref.~\cite{Altmannshofer:2008dz}, in which
\begin{eqnarray}\label{O910}
O_9=\frac{e^2}{g_s^2}(\bar{d}\gamma_\mu P_Lb)(\bar{l}\gamma^\mu l)\,,\quad O_{10}=\frac{e^2}{g_s^2}(\bar{d}\gamma_\mu P_Lb)(\bar{l}\gamma^\mu\gamma_5 l)\,.
\end{eqnarray}
Wilson coefficients $C_i$ can be calculated perturbatively~\cite{Beneke:2001at,bobeth,bobeth02,Huber:2005ig}, with the numerical results listed in Table~\ref{wc}.
The effective coefficients $C_{7,9}^{eff}$, which are particular combinations of $C_{7,9}$ with the other $C_i$, are defined as~\cite{Altmannshofer:2008dz}
\begin{eqnarray}\label{eq:effWC}
&& C_7^{\rm eff} = \frac{4\pi}{\alpha_s}\, C_7 -\frac{1}{3}\, C_3 -
\frac{4}{9}\, C_4 - \frac{20}{3}\, C_5\, -\frac{80}{9}\,C_6\,,
\nonumber\\
&& C_9^{\rm eff} = \frac{4\pi}{\alpha_s}\,C_9 + Y(q^2)\,, \qquad
   C_{10}^{\rm eff} = \frac{4\pi}{\alpha_s}\,C_{10}\,,
\end{eqnarray}
in which $Y(q^2)$ denotes the matrix element of four-quark operators and given by
\begin{eqnarray}
Y(q^2)&=&h(q^2,m_c)\big(\frac{4}{3}C_1+C_2+6C_3+60C_5\big)-\frac{1}{2}h(q^2,m_b)\big(7C_3+\frac{4}{3}C_4+76C_5+\frac{64}{3}C_6\big)\,\nonumber\\
&&-\frac{1}{2}h(q^2,0)\big(C_3+\frac{4}{3}C_4+16C_5+\frac{64}{3}C_6\big)+\frac{4}{3}C_3+\frac{64}{9}C_5+\frac{64}{27}C_6\,.
\end{eqnarray}
We have neglected the long-distance contribution mainly due to $J/\Psi$ and $\Psi^{\prime}$ in the decay chain $\bar{B}_s \to \phi \Psi^{(\prime)}\to\phi l^+l^-$,
which could be vetoed experimentally~\cite{CDFPhill}. For the recent detailed discussion of such resonance effects, we refer to Ref.~\cite{resEff}.
\begin{table}[htbp]
 \begin{center}
 \caption{The SM Wilson coefficients at the scale $\mu=m_b$.}
 \label{wc}
 \vspace{0.3cm}
 \doublerulesep 0.5pt \tabcolsep 0.07in
 \begin{tabular}{lccccccccccc}
 \hline \hline
 $C_1(m_b)$& $C_2(m_b)$& $C_3(m_b)$& $C_4(m_b)$& $C_5(m_b)$& $C_6(m_b)$& $C_7^{\rm eff}(m_b)$& $C_9^{\rm eff}(m_b)-Y(q^2)$& $C_{10}^{\rm eff}(m_b)$\\\hline
 $-0.284$  & $1.007$   & $-0.004$  & $-0.078$   & $0.000$   & $0.001$   & $-0.303$            & $4.095$   & $-4.153$\\
  \hline \hline
 \end{tabular}
 \end{center}
 \end{table}

Although there are quite a lot of interesting observables in semileptonic $b\to s \ell^+ \ell^-$ decay, we shall focus only on the dilepton invariant mass spectrum and the forward-backward asymmetry in this paper. Adopting the same convention and notation as \cite{Ali:1999mm}, the dilepton invariant mass spectrum and forward-backward asymmetry for $\bar{B}_s\to \phi\ell^+\ell^-$ decay is given as
\begin{eqnarray}
\frac{\d \gl^{\phi}}{\d\sh} & = &
  \frac{G_F^2 \, \alpha^2 \, m_{B_s}^5}{2^{10} \pi^5}
      \left| V_{ts}^\ast  V_{tb} \right|^2 \, \uh(\sh) \,
      \Bigg\{
\frac{|A|^2}{3} \sh \la (1+2 \frac{\mlh^2}{\sh}) +|E|^2 \sh
\frac{\uh(\sh)^2}{3}
        \Bigg.
        \nonumber \\
  & & + \Bigg. \frac{1}{4 \mvh^2} \left[
|B|^2 (\la-\frac{\uh(\sh)^2}{3} + 8 \mvh^2 (\sh+ 2 \mlh^2) )
          + |F|^2 (\la -\frac{ \uh(\sh)^2}{3} + 8 \mvh^2 (\sh- 4 \mlh^2))
\right]
        \Bigg.
        \nonumber \\
  & & +\Bigg.
   \frac{\la }{4 \mvh^2} \left[ |C|^2 (\la - \frac{\uh(\sh)^2}{3})
 + |G|^2 \left(\la -\frac{\uh(\sh)^2}{3}+4 \mlh^2(2+2 \mvh^2-\sh) \right)
\right]
        \Bigg.
        \nonumber \\
  & & - \Bigg.
   \frac{1}{2 \mvh^2}
\left[ {\rm Re}(BC^\ast) (\la -\frac{ \uh(\sh)^2}{3})(1 - \mvh^2 -
\sh)
\nonumber  \right. \Bigg.\\
& & + \left.  \Bigg.
       {\rm Re}(FG^\ast) ((\la -\frac{ \uh(\sh)^2}{3})(1 - \mvh^2 - \sh) +
4 \mlh^2 \la) \right]
        \Bigg.
        \nonumber \\
  & & - \Bigg.
 2 \frac{\mlh^2}{\mvh^2} \la  \left[ {\rm Re}(FH^\ast)-
 {\rm Re}(GH^\ast) (1-\mvh^2) \right] +\frac{\mlh^2}{\mvh^2} \sh \la |H|^2
  \Bigg\} \,;
   \label{eq:dwbvll}
\end{eqnarray}
\begin{eqnarray}\label{EqAFB}
  \frac{\d \a_{\rm FB}^{\phi}}{\d \sh}& =&
  -\frac{G_F^2 \, \alpha^2 \, m_{B_s}^5}{2^{8} \pi^5}
      \left| V_{ts}^\ast  V_{tb} \right|^2 \, \sh \, \uh(\sh)^2 \, \nonumber \\
& & \hspace{-0.5cm} \times \left[  {\rm Re}(\cne\ct^{\ast}) V A_1+ \frac{\mbh}{\sh}
{\rm Re}(\cse\ct^{\ast}) {\Big (}V T_2 (1-\mvh)+ A_1 T_1
(1+\mvh){\Big)} \right]\,,
\end{eqnarray}
with $s=q^2$ and $\hat{s}=s/m_{B_s}^2$. Here the auxiliary functions $A\,,B\,,C\,,E\,,F$ and $G$, with the explicit expressions given in Ref.~\cite{Ali:1999mm}, are combinations of the effective Wilson coefficients in Eq.~(\ref{eq:effWC}) and the $B_s\to\phi$ transition form factors, which are calculated with light-cone QCD sum rule approach in Ref.~\cite{Ball:2004rg} and given in Appendix B.
From the experimental point of view, the normalized forward-backward asymmetry is more useful, which is defined
as~\cite{Ali:1999mm}
\begin{equation} \label{eq:AFB2}
  \frac{\d \bar{\a}_{\rm FB}}{\d \sh} =
\frac{\d \a_{\rm FB}}{\d \sh}/\frac{\d \gl}{\d\sh}\,.
\end{equation}

A new family non-universal $Z^\prime$ boson could be naturally derived in many extension of SM. One of the possible way to get such non-universal $Z^\prime$ boson is to include an addition $U^{\prime}(1)$ gauge symmetry, which has been formulated in detail by Langacker and Pl\"{u}macher~\cite{Langacker}. Under the assumption that the couplings of right-handed quark flavors with $Z^{\prime}$ boson are diagonal, the $Z^{\prime}$ part of the effective Hamiltonian for $b\to s l^+ l^-$ transition can be written as~\cite{Liu}
\begin{equation}\label{ZPHbsll}
 {\cal H}_{eff}^{Z^{\prime}}(b\to sl^+l^-)=-\frac{2G_F}{\sqrt{2}}
 V_{tb}V^{\ast}_{ts}\Big[-\frac{B_{sb}^{L}B_{ll}^{L}}{V_{tb}V^{\ast}_{ts}}
 (\bar{s}b)_{V-A}(\bar{l}l)_{V-A}-\frac{B_{sb}^{L}B_{ll}^{R}}{V_{tb}V^{\ast}_{ts}}
 (\bar{s}b)_{V-A}(\bar{l}l)_{V+A}\Big]+{\rm h.c.}\,.
\end{equation}
With the assumption that no significant RG running effect between $M_{Z^{\prime}}$ and $M_W$ scales, $Z^{\prime}$ contributions could be treated as modification to wilson coefficients, i.e. $C_{9,10}^{\prime}(M_W)=C_{9,10}^{SM}(M_W)+\triangle C_9^{\prime}(M_W)$. As a result, Eq.~(\ref{ZPHbsll}) could also be reformulated as
\begin{equation}\label{ZPHbsllMo}
 {\cal H}_{eff}^{Z^{\prime}}(b\to sl^+l^-)=-\frac{4G_F}{\sqrt{2}}
 V_{tb}V^{\ast}_{ts}\left[\triangle C_9^{\prime} O_9+\triangle C_{10}^{\prime} O_{10}\right]+{\rm h.c.}\,,
\end{equation}
with
\begin{eqnarray}\label{C910Zp}
 \triangle C_9^{\prime}(M_W)&=&-\frac{g_s^2}{e^2}\frac{B_{sb}^L
 }{V_{ts}^{\ast}V_{tb}} S_{ll}^{LR}\,,\quad  S_{ll}^{LR}=(B_{ll}^{L}+B_{ll}^{R})\,,\nonumber \\
 \triangle C_{10}^{\prime}(M_W)&=&\frac{g_s^2}{e^2}\frac{B_{sb}^L
 }{V_{ts}^{\ast}V_{tb}} D_{ll}^{LR}\,,\quad\,\,  D_{ll}^{LR}=(B_{ll}^{L}-B_{ll}^{R})\,.
\end{eqnarray}
$B_{sb}^L$ and $B_{ll}^{L,R}$ denote the effective chiral $Z^{\prime}$ couplings to quarks and leptons, in which the off-diagonal element $B_{sb}^L$ can contain a new weak phase and could be written as  $|B_{sb}^L|e^{i\phi_s^{L}}$.

To include $Z^{\prime}$ contributions, one just needs to make the replacements
\begin{eqnarray}\label{C910ERp}
C_{9}^{\rm eff}&\to&\bar{C}_9^{\rm eff}=\frac{4\pi}{\alpha_s}C_9^{\prime}+Y(q^2)\;,\nonumber\\
C_{10}^{\rm eff}&\to&\bar{C}_{10}^{\rm eff}=\frac{4\pi}{\alpha_s}C_{10}^{\prime}\;,
\end{eqnarray}
in the formalisms relevant to $\bar{B}_s\to \phi \ell^{+}\ell^{-}$.

\section{Numerical analyses and discussions}
\begin{table}[t]
 \begin{center}
 \caption{Predictions for ${\cal B}(\bar{B}_s\to \phi \mu^+\mu^-)[\times 10^{-6}]$ and $A_{FB}(\bar{B}_s\to \phi \mu^+\mu^-)[\times 10^{-2}]$ within the SM and the non-universal $Z^{\prime}$ model.}
 \label{tab_pred}
 \vspace{0.5cm}
 \small
 \doublerulesep 0.7pt \tabcolsep 0.1in
 \begin{tabular}{lccccccccccc} \hline \hline
             &Exp.~\cite{HFAG} & SM           & S1           &S2            &Scen.~I &Scen.~II & Scen.~III
 \\\hline
 ${\cal B}$  &$1.44\pm0.57$    &$1.46\pm0.10$ &$2.47\pm1.18$ &$1.40\pm0.27$ &$2.86$  &$1.26$   &$1.92$ \\
 ${\cal B}^L$&---              &$0.34\pm0.04$ &$0.56\pm0.27$ &$2.61\pm0.19$ &$0.64$  &$0.28$   &$0.44$\\
 ${\cal B}^H$&---              &$0.29\pm0.02$ &$0.51\pm0.25$ &$1.26\pm0.08$ &$0.59$  &$0.26$   &$0.39$\\
 \hline
 $A_{FB}$    &---              &$25.6\pm1.2$  &$19.4\pm10.9$ &$24\pm0.03$   &$29.9$  &$26.6$   &$8.9$\\
 $A_{FB}^L$  &---              &$5.7\pm0.6$   &$6.0\pm7.4$   &$0.09\pm0.02$ &$13.3$  &$6.9$    &$1.4$\\
 $A_{FB}^H$  &---              &$34.1\pm0.2$  &$22.5\pm12.9$ &$-0.07\pm0.01$&$35.0$  &$34.8$   &$13.1$\\
  \hline \hline
 \end{tabular}
 \end{center}
 \end{table}
\begin{figure}[ht]
\begin{center}
\epsfxsize=15cm \centerline{\epsffile{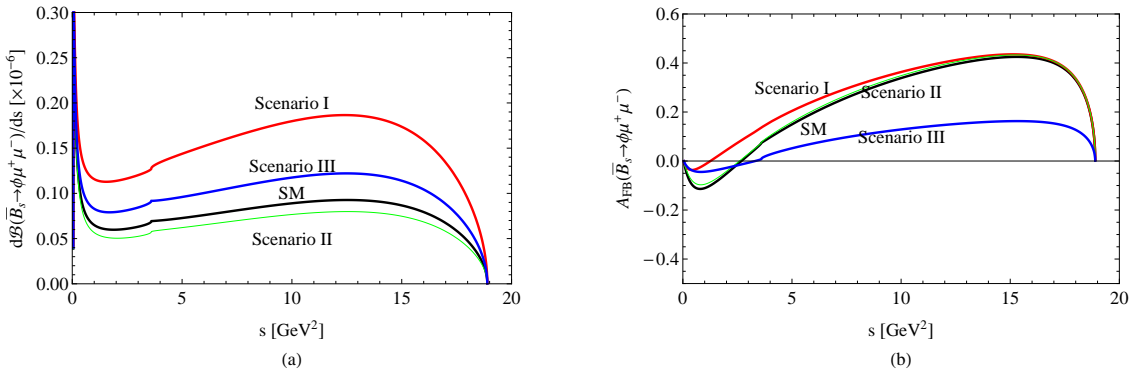}}
\centerline{\parbox{16cm}{\caption{\label{spectr}\small Dimuon invariant mass distribution and normalized forward-backward asymmetry of the $\bar{B}_s\to \phi \mu^+\mu^-$ decay within SM and three limiting scenarios.}}}
\end{center}
\end{figure}

\begin{table}[t]
 \begin{center}
 \caption{The inputs parameters for the $Z^{\prime}$ couplings~\cite{Chang2,Chang3}. }
 \label{NPPara_value}
 \vspace{0.5cm}
 \doublerulesep 0.7pt \tabcolsep 0.1in
 \begin{tabular}{lccccccccccc} \hline \hline
    & $|B_{sb}^L|(\times10^{-3})$ & $\phi_{s}^L[^{\circ}]$ &$S^{LR}_{\mu\mu}(\times10^{-2})$ & $D^{LR}_{\mu\mu}(\times10^{-2})$\\\hline
 S1 & $1.09\pm0.22$               & $-72\pm7$              &$-2.8\pm3.9$                     & $-6.7\pm2.6$ \\
 S2 & $2.20\pm0.15$               & $-82\pm4$              &$-1.2\pm1.4$                     & $-2.5\pm0.9$ \\
  \hline \hline
 \end{tabular}
 \end{center}
 \end{table}
With the relevant theoretical formulas collected in
Section~\ref{theo} and the input parameters summarized in the
Appendix, we now proceed to present our numerical analyses and
discussions.

In Table~\ref{tab_pred}, we present our theoretical predictions for integrated branching fraction and
forward-backward asymmetry of $\bar{B}_s\to \phi \mu^+\mu^-$ decay. Within the SM, we again find our prediction ${\cal B}^{SM}(\bar{B}_s\to \phi \mu^+\mu^-)=1.46\times 10^{-6}$ is perfectly consistent with CDF measurement $(1.44\pm0.57)\times 10^{-6}$. The forward-backward asymmetry for $\bar{B}_s\to \phi \mu^+\mu^-$ decay is evaluated at $\sim25\%$,  which hasn't be measured by the experiment. In addition, in Table~\ref{tab_pred}, we also calculate their results ${\cal B}^{L,H}$ and $A_{FB}^{L,H}$ at both low~($1GeV^2<s<6GeV^2$) and high~($14.4GeV^2<s<25GeV^2$) integration regions, which are sufficiently below and above the threshold for charmonium resonances $J/\psi,\psi^{\prime}$ respectively. The dimuon invariant mass distribution and
forward-backward asymmetry spectrum are shown in Fig.~\ref{spectr}. As Fig.~\ref{spectr}(b) shows, similar to the situation in $\bar{B}^0\to K^{\ast}\mu^+\mu^-$ decay, the zero crossing exist in $A_{FB}$ spectrum at $s_0\sim3~GeV^2$, whose position is well-determined and free from hadronic uncertainties at the leading order in $\alpha_s$~\cite{Beneke:2001at,Ali:1999mm,Burdman:1998mk}. In $\bar{B}^0\to K^{\ast}\mu^+\mu^-$ decay, the $A_{FB}$ spectrum measured by Belle collaboration~\cite{:2009zv} indicates that there might be no zero crossing, which presents a challenge to the SM in low $s$ region.
If the future measurement on $A_{FB}(\bar{B}_s\to \phi\mu^+\mu^-)$ spectrum presents a similar result as the one in $\bar{B}^0\to K^{\ast}\mu^+\mu^-$ decay, it will be a significant NP signal.

Within a family non-universal $Z^{\prime}$ model, the $Z^{\prime}$ contributions to $\bar{B}_s\to \phi \mu^+\mu^-$ decay involve
four new $Z^{\prime}$ parameters $|B_{sb}^L|$, $\phi_{s}^L$, $S^{LR}_{\mu\mu}$ and $D^{LR}_{\mu\mu}$. Combining the constraints from $\bar{B}_s-B_s$ mixing, $B\to\pi K^{(\ast)}$ and $\rho K$ decays, $|B_{sb}^L|$ and $\phi_{s}^L$ have been strictly constrained~\cite{Chang1,Chang2}. After having included the constraints from $\bar{B}_d\to X_s\mu\mu$, $K\mu\mu$ and $K^{\ast}\mu\mu$, as well as $B_s\to\mu\mu$ decays, we have also gotten the allowed ranges for $S^{LR}_{\mu\mu}$ and $D^{LR}_{\mu\mu}$ in Ref.~\cite{Chang3}.
For convenience, we recollect their numerical results in Table~\ref{NPPara_value}, in which S1 and S2 correspond to UTfit collaboration's two fitting results for $\bar{B}_s-B_s$ mixing~\cite{UTfit}. Our following evaluations and discussions are based on these given ranges for $Z^{\prime}$ couplings. With the values of $Z^{\prime}$ parameters  listed in Table~\ref{NPPara_value} as inputs, we present our predictions for the observables in the third and fourth columns of Table~\ref{tab_pred}.

\begin{figure}[t]
\begin{center}
\epsfxsize=15cm \centerline{\epsffile{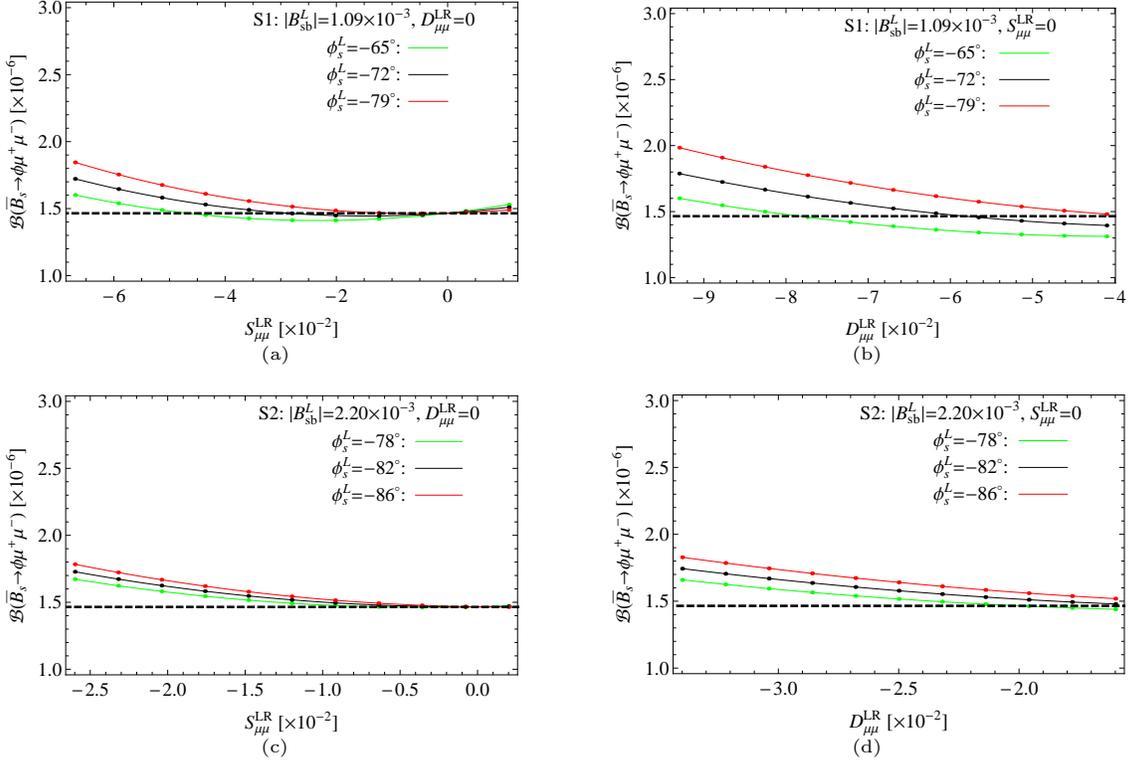}}
\centerline{\parbox{16cm}{\caption{\label{Br_ZpCoupling}\small The dependence of ${\cal B}(\bar{B}_s\to \phi \mu^+\mu^-)$ on $S_{\mu\mu}^{LR}$ and $D_{\mu\mu}^{LR}$ within their allowed ranges in S1 and S2 with different $\phi_s^L$ values. The black dashed line corresponds to the SM result.}}}
\end{center}
\end{figure}
\begin{figure}[ht]
\begin{center}
\epsfxsize=15cm \centerline{\epsffile{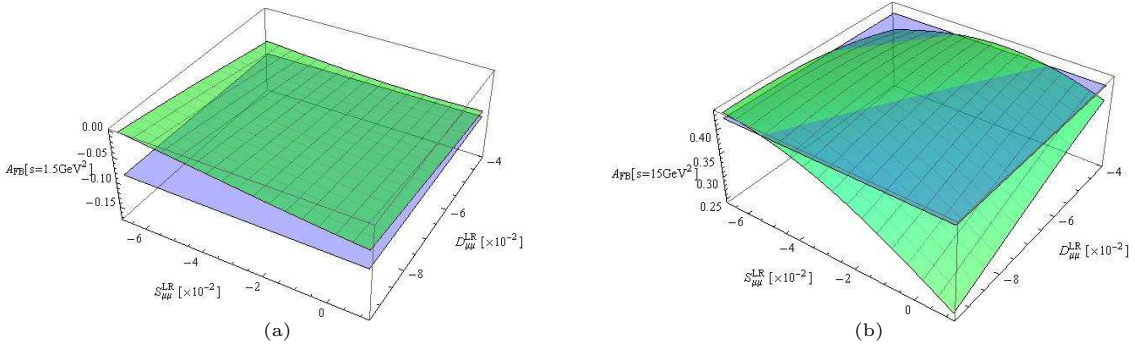}}
\centerline{\parbox{16cm}{\caption{\label{AFBLHZP}\small The dependence of $A_{FB}(\bar{B}_s\to \phi\mu^+\mu^-)$ on $S_{ud}^{L,R}$ and $D_{ud}^{L,R}$ at $s=1.5\,{\rm GeV^2}$~(a) and $s=15\,{\rm GeV^2}$~(b) with $|B_{db}^L|=1.09(\times10^{-3})$, $\phi_s^L=-72^{\circ}$~(S1) and the central values of the other theoretical input parameters. The blue planes correspond to SM results.}}}
\end{center}
\end{figure}
As illustrated in Fig.~\ref{Br_ZpCoupling}, integrated branching fraction for $\bar{B}_s\to \phi \mu^+\mu^-$ is sensitive to the $Z^{\prime}$ contributions. Obviously, $\bar{B}_s\to \phi \mu^+\mu^-$ is enhanced by the $Z^{\prime}$ contributions with large negative $S_{\mu\mu}^{LR}$, $D_{\mu\mu}^{LR}$ and $\phi_s^L$. Moreover, compared Fig.~\ref{Br_ZpCoupling}~(a,b) with (c,d), we find the effects of solution S1 is more significant than the one of S2. So,  for simplicity, we just pay our attention to the solution S1 in the following. As Fig.~\ref{Br_ZpCoupling} shows, the $Z^{\prime}$ contributions with a small negative weak phase $\phi_s^L$ are helpful to reduce ${\cal B}(\bar{B}_s\to \phi \mu^+\mu^-)$. However, because the range $\phi_s^L>-65^{\circ}$ is excluded by the constraints from  $\bar{B}_s-B_s$ mixing and $B\to \pi K$ decays~\cite{Chang1,Chang2}, ${\cal B}(\bar{B}_s\to \phi \mu^+\mu^-)$ is hardly to be reduced so much by $Z^{\prime}$ contributions.

In order to see the $Z^{\prime}$ effect on $A_{FB}(\bar{B}_s\to \phi \mu^+\mu^-)$ explicitly, with $Y(q^2)$ being excluded, we can
rewrite ${\rm Re}(\bar{C}_9^{\rm eff}\bar{C}_{10}^{{\rm eff}\ast})$ and ${\rm Re}(\bar{C}_7^{\rm eff}\bar{C}_{10}^{{\rm eff}\ast})$ in Eq.~(\ref{EqAFB}), which dominates $A_{FB}(\bar{B}_s\to \phi \mu^+\mu^-)$ in high and low $s$ regions respectively, as
\begin{eqnarray}\label{Ldomi}
{\rm Re}(\bar{C}_9^{\rm eff}\bar{C}_{10}^{{\rm eff}\ast})&\simeq&{\rm Re}(\bar{C}_9^{\rm eff})\,{\rm Re}(\bar{C}_{10}^{{\rm eff}\ast})+\left(\frac{4\pi}{\alpha_s}\right)^2\,{\rm Im}(\triangle C_9^{\prime})\,{\rm Im}(\triangle C_{10}^{\prime})\,,\\\label{Hdomi}
{\rm Re}(\bar{C}_7^{\rm eff}\bar{C}_{10}^{{\rm eff}\ast})&\simeq&{\rm Re}(\bar{C}_7^{\rm eff}){\rm Re}(\bar{C}_{10}^{{\rm eff}\ast})\,.
\end{eqnarray}
Combining Eq.~(\ref{C910Zp}) and Eq.~(\ref{Hdomi}), due to the tiny $Z^{\prime}$ contribution to $C_7^{\rm eff}$, the only solution to enhance $A_{FB}$ in low $s$ region is a larger negative $D^{LR}_{\mu\mu}$, which also can be found in Fig.~\ref{AFBLHZP}(a). In high $s$ region, as  Fig.~\ref{AFBLHZP}~(b) shows, $A_{FB}$ could be reduced significantly and enhanced a bit by $Z^{\prime}$ contributions.

Based on the  analyses above, in order to evaluate the exact strength of $Z^{\prime}$ effects, our following analyses can be divided into three limiting scenarios:

\subsubsection*{\small Scenario~I}
In order to get the maximum ${\cal B}(\bar{B}_s\to \phi \mu^+\mu^-)$, within the allowed ranges for $Z^{\prime}$ couplings listed in Table~\ref{NPPara_value}, we choose a set of extreme values
\begin{eqnarray}\label{ScenarioI}
|B_{sb}^L|=1.31\times10^{-3}\,, \phi_{s}^L=-79^{\circ}\,, S^{LR}_{\mu\mu}= -6.7\times10^{-2}\,, D^{LR}_{\mu\mu}=-9.3\times10^{-2} \quad {\rm Scen.~I}\,,
\end{eqnarray}
named Scenario~I. With the central values of the other theoretical input parameters, we get ${\cal B}(\bar{B}_s\to \phi \mu^+\mu^-)=2.86\times10^{-6}$, which is $2.5\sigma$ larger than CDF result $(1.44\pm0.57)\times10^{-6}$. Compared with the SM prediction $1.46\times10^{-6}$, we find ${\cal B}(\bar{B}_s\to \phi \mu^+\mu^-)$ could be enhanced by about $96\%$ at most by $Z^{\prime}$ contributions.

This scenario is the most helpful solution to moderate the discrepancy for $A_{FB}(\bar{B}_s\to K^{\ast} \mu^+\mu^-)$ between SM prediction and experimental data in low $s$ region~\cite{Chang3,CDLv}. As Fig.~\ref{AFBLHZP}~(a) shows, we find Scenario~I also provides the most helpful solution to enhance $A_{FB}(\bar{B}_s\to \phi \mu^+\mu^-)$ in low $s$ region. Compared with the SM results, we find $A_{FB}^{(L)}(\bar{B}_s\to \phi \mu^+\mu^-)$ could be enhanced by about $17\%(133.3\%)$ at most. However,
in the high $s$ region, the effect of Scenario~I on $A_{FB}(\bar{B}_s\to \phi \mu^+\mu^-)$, as Fig.~\ref{AFBLHZP}~(b) shows, is not significant.

In addition, due to the strong constraints on $D^{LR}_{\mu\mu}$ from $\bar{B}_d\to X_s\mu\mu$ decay, the much larger value $|D^{LR}_{\mu\mu}|>9.3\times10^{-2}$ is forbidden~\cite{Chang2}, which means the sign of ${\rm Re}(\bar{C}_7^{\rm eff}\bar{C}_{10}^{{\rm eff}\ast})$ can hardly be flipped by $Z^{\prime}$ contributions~\cite{Chang3}. So, as Fig.~\ref{spectr}~(b) shows, the zero crossing in  $A_{FB}$ spectrum also exists and moves to $s_0\sim1GeV^2$ point in this scenario.

\subsubsection*{\small Scenario~II}
From Fig.~\ref{Br_ZpCoupling}, one may find that ${\cal B}(\bar{B}_s\to \phi \mu^+\mu^-)$ can hardly be reduced by $Z^{\prime}$ contributions so much within the allowed $Z^{\prime}$ parameters' ranges. The most minimal value of ${\cal B}(\bar{B}_s\to \phi \mu^+\mu^-)$ appears at
\begin{eqnarray}\label{ScenarioII}
|B_{sb}^L|=1.31\times10^{-3}\,, \phi_{s}^L=-65^{\circ}\,, S^{LR}_{\mu\mu}= -2\times10^{-2}\,, D^{LR}_{\mu\mu}=-4\times10^{-2} \quad {\rm Scen.~II}\,,
\end{eqnarray}
named Scenario~II. In this scenario, compared with SM prediction, we find ${\cal B}(\bar{B}_s\to \phi \mu^+\mu^-)$ could be reduced just by about $14\%$ at most by $Z^{\prime}$ contributions. Due to the small $Z^{\prime}$ contributions, its effect on $A_{FB}(\bar{B}_s\to \phi \mu^+\mu^-)$ is also tiny.

\subsubsection*{\small Scenario~III}
As Fig.~\ref{AFBLHZP}~(b) shows, $A_{FB}(\bar{B}_s\to \phi \mu^+\mu^-)$ would be reduced rapidly in high $s$ region when $S^{LR}_{\mu\mu}$ is enlarged. So, we present a limiting scenario for the minimal $A_{FB}^H(\bar{B}_s\to \phi \mu^+\mu^-)$,
\begin{eqnarray}\label{ScenarioIII}
|B_{sb}^L|=1.31\times10^{-3}\,, \phi_{s}^L=-65^{\circ}\,, S^{LR}_{\mu\mu}= 1.1\times10^{-2}\,, D^{LR}_{\mu\mu}=-9.3\times10^{-2} \quad {\rm Scen.~III}\,,
\end{eqnarray}
named Scenario~III. Compared with SM prediction, $A_{FB}^{(H)}(\bar{B}_s\to \phi \mu^+\mu^-)$ is reduced by about $62\%$~($62\%$). However, as Fig.~\ref{AFBLHZP}~(b) shows, in the low $s$ region, $A_{FB}$ is just enhanced a bit. So, this scenario also leads to the minimal $A_{FB}(\bar{B}_s\to \phi \mu^+\mu^-)\sim8.9\%$, which is $65\%$ smaller than SM prediction. While, in this scenario, our prediction ${\cal B}(\bar{B}_s\to \phi \mu^+\mu^-)=1.92\times10^{-6}$ also agrees with CDF measurement within $1\sigma$. So, although Scenario~III presents a strange effects on $A_{FB}$ spectrum, it is not excluded by current measurement either.
Moreover, different from Scenario~I, zero crossing in $A_{FB}$ spectrum moves to positive side in this scenario.

\section{Conclusion}
In conclusion, motivated by recent measurement on ${\cal B}(\bar{B}_s\to \phi \mu^+\mu^-)$ by CDF Collaboration, after revisiting $\bar{B}_s\to \phi \mu^+\mu^-$ decay within SM, we have investigated the effects of a family non-universal $Z^{\prime}$ boson with the given $Z^{\prime}$ couplings. Our conclusions can be summarized as:
\begin{itemize}
\item Branching fraction and forward-backward asymmetry for $\bar{B}_s\to \phi \mu^+\mu^-$ decay are sensitive to $Z^{\prime}$ contributions. All of the $Z^{\prime}$ couplings listed in Table~\ref{NPPara_value} survive under the constraint from ${\cal B}(\bar{B}_s\to \phi \mu^+\mu^-)$ measured by CDF within errors.

\item We present three limiting scenarios: ${\cal B}(\bar{B}_s\to \pi^- K^+)$ and $A_{FB}^{(L)}(\bar{B}_s\to \phi \mu^+\mu^-)$ could be enhanced by about $96\%$ and $17\%\,(133\%)$ at most by $Z^{\prime}$ contributions (Scenario~I); However, ${\cal B}(\bar{B}_s\to \pi^- K^+)$ is hardly to be reduced ( reduced by $14\%$ at most in Scenario~II) by $Z^{\prime}$ contributions; Moreover, in Scenario~III, $A_{FB}^{(H)}(\bar{B}_s\to \phi \mu^+\mu^-)$ reaches its minimal value, which is $65\%(62\%)$ lower than SM prediction.

\item The zero crossing in $A_{FB}(\bar{B}_s\to \phi \mu^+\mu^-)$ spectrum always exists in the three scenarios.
\end{itemize}

The refined measurements for the $B_{s}$ leptonic
decay $\bar{B}_s\to \phi \mu^+\mu^-$  in the upcoming LHC-b and proposed super-B
will provide a powerful  testing ground for the SM and possible NP scenarios. Our analyses
of the $Z^{\prime}$ effects on the observables for $\bar{B}_s\to \phi \mu^+\mu^-$ decay are useful for
probing or refuting the effects of a family non-universal $Z^{\prime}$ boson.

\section*{Acknowledgments}
The work is supported by the National Science Foundation
under contract Nos.11075059, 10735080 and 11005032.

\begin{appendix}

\section*{Appendix A: Theoretical input parameters}

For the CKM matrix elements, we adopt the UTfit collaboration's
fitting results~\cite{UTfitCKM}
\begin{eqnarray}
\overline{\rho}&=&0.132\pm0.02\,(0.135\pm0.04), \quad
\overline{\eta}=0.367\pm0.013\,(0.374\pm0.026),\nonumber\\
A&=&0.8095\pm0.0095\,(0.804\pm0.01),\quad
\lambda=0.22545\pm0.00065\,(0.22535\pm0.00065).
\end{eqnarray}

As for the quark masses, we take~\cite{PDG10,PMass}
\begin{eqnarray}
&&m_u=m_d=m_s=0, \quad m_c=1.61^{+0.08}_{-0.12}\,{\rm GeV},\nonumber\\
&&m_b=4.79^{+0.19}_{-0.08}\,{\rm GeV}, \quad m_t=172.4\pm1.22\,{\rm GeV}.
\end{eqnarray}

\section*{Appendix B: Transition form factors from light-cone QCD sum rule}

In order to calculate the $\bar{B}_s\to\phi \ell^+ \ell^-$ decay amplitude, we have to evaluate the $\bar{B}_s\to\phi$ matrix elements of quark bilinear currents.
They can be expressed in terms of ten form factors, which depend on the momentum transfer $q^2$ between the $B_s$ and the $\phi$ mesons~($q=p - k$)~\cite{Ball:2004rg}:

\begin{eqnarray}\label{eq:SLFF}
\langle  \phi(k) | \bar d\gamma_\mu(1-\gamma_5) b | \bar
B_s(p)\rangle  &=& -i \epsilon^*_\mu (m_{B_s}+m_{\phi}) A_1(q^2) + i
(2p-q)_\mu (\epsilon^* \cdot q)\,
\frac{A_2(q^2)}{m_{B_s}+m_{\phi}}\, \nonumber \\
& & + i q_\mu (\epsilon^* \cdot q) \, \frac{2m_{\phi}}{q^2}\,
\Big[A_3(q^2)-A_0(q^2)\Big] \, \nonumber \\
& & + \epsilon_{\mu\nu\rho\sigma}\epsilon^{*\nu} p^\rho k^\sigma\,
\frac{2V(q^2)}{m_{B_s}+m_{\phi}}\,,
\end{eqnarray}
with $A_3(q^2) = \frac{m_{B_s}+m_{\phi}}{2m_{\phi}}\, A_1(q^2) -
\frac{m_{B_s}-m_{\phi}}{2m_{\phi}}\, A_2(q^2)$ and $A_0(0) =  A_3(0)$,
\begin{eqnarray}\label{eq:pengFF}
\langle \phi(k) | \bar s \sigma_{\mu\nu} q^\nu (1+\gamma_5) b |
\bar{B}_s(p)\rangle &=& i\epsilon_{\mu\nu\rho\sigma} \epsilon^{*\nu}
p^\rho k^\sigma \, 2 T_1(q^2)\, \nonumber\\
& & + T_2(q^2) \Big[\epsilon^*_\mu (m_{B_s}^2-m_{\phi}^2)
- (\epsilon^* \cdot q) \,(2p-q)_\mu \Big] \, \nonumber\\
& & + T_3(q^2) (\epsilon^* \cdot q) \left[q_\mu -
\frac{q^2}{m_{B_s}^2-m_{\phi}^2}\, (2p-q)_\mu \right]\,,
\end{eqnarray}
with $T_1(0) = T_2(0)$. $\epsilon_\mu$ is the polarization vector of the $\phi$ meson. The physical range in $s=q^2$ extends from $s_{\rm min} = 0$ to $s_{\rm max} =(m_{B_s}-m_{\phi})^2$.

\begin{table}[t]
\begin{center}
\caption{\label{FFfit} Fit parameters for $B_s\to\phi$ transition form
factors~\cite{Ball:2004rg}.}
\vspace{0.3cm}
\begin{tabular}{crrrrrl}\hline\hline
                         &$F(0)$   &$r_1$   &$m_R^2$   &$r_2$     &$m^2_{\rm fit}$& \\ \hline
$V^{B_s\to\phi}$         &$0.434$  &$1.484$ &$5.32^2$  &$-1.049$  &$39.52$         &Eq.~(\ref{r12mRfit})\\\hline
$A_0^{B_s\to\phi}$       &$0.474$  &$3.310$ &$5.28^2$  &$-2.835$  &$31.57$         &Eq.~(\ref{r12mRfit})\\\hline
$A_1^{B_s\to\phi}$       &$0.311$  &---     &---       &$0.308$   &$36.54$         &Eq.~(\ref{r2mfit})\\\hline
$A_2^{B_s\to\phi}$       &$0.234$  &$-0.054$&---       &$0.288$   &$48.94$         &Eq.~(\ref{r12mfit})\\\hline
$T_1^{B_s\to\phi}$       &$0.349$  &$1.303$ &$5.32^2$  &$-0.954$  &$38.28$         &Eq.~(\ref{r12mRfit})\\\hline
$T_2^{B_s\to\phi}$       &$0.349$  &---     &---       &$0.349$   &$37.21$         &Eq.~(\ref{r2mfit})\\\hline
$\tilde{T}_3^{B\to \phi}$&$0.349$  &$0.027$ &---       &$0.321$   &$45.56$         &Eq.~(\ref{r12mfit})\\
\hline\hline
\end{tabular}
\end{center}
\end{table}
These transition form factors have been updated recently within the light-cone QCD sum rule approach~\cite{Ball:2004rg}. For the $q^2$ dependence of the form factors, they can be parameterized in terms of simple formulae with two or three parameters. The form factors $V$, $A_0$ and $T_1$ are parameterized by
\begin{eqnarray}
F(s)=\frac{r_1}{1-s/m^2_{R}}+\frac{r_2}{1-s/m^2_{\rm fit}}.
\label{r12mRfit}
\end{eqnarray}
For the form factors $A_2$ and $\tilde{T}_3$, it is more appropriate to expand to the second order around the pole, yielding
\begin{eqnarray}
F(s)=\frac{r_1}{1-s/m^2}+\frac{r_2}{(1-s/m)^2}\,, \label{r12mfit}
\end{eqnarray}
where $m=m_{\rm fit}$ for $A_2$ and $\tilde{T}_3$. The fit formula for $A_1$ and $T_2$ is
\begin{eqnarray}
F(s)=\frac{r_2}{1-s/m^2_{\rm fit}}.\label{r2mfit}
\end{eqnarray}
The form factor $T_3$ can be obtained through the relation $T_3(s)=\frac{m_{B_s}^2-m_{\phi}^2}{s}\big[\tilde{T}_3(s)-T_2(s)\big]$. All the relevant fitting parameters for these form factors are taken from Ref.~\cite{Ball:2004rg} and are recollected in Table~\ref{FFfit}.

\end{appendix}

\end{document}